\documentclass[twocolumn,superscriptaddress,prl,apsrev4-1.bst]{revtex4}
\usepackage{inputenc}
\usepackage[english]{babel}
\usepackage{graphicx}
\usepackage{amsmath}
\usepackage{amssymb}
\usepackage{color}
\usepackage{textcomp}
\usepackage{lmodern}
\usepackage[TS1,T1]{fontenc}
\usepackage[amssymb]{SIunits}
\usepackage{multirow}

\begin{document}

\title{Interactions and Charge Fractionalization in an Electronic\\
Hong-Ou-Mandel Interferometer}
\keywords{Electronic quantum optics}

\author{Claire Wahl}
\email{wahl@cpt.univ-mrs.fr}
\affiliation{Aix Marseille Universit\'e, CNRS, CPT, UMR 7332, 13288 Marseille, France}
\affiliation{Universit\'e de Toulon, CNRS, CPT, UMR 7332, 83957 La Garde, France}
\author{J\'er\^ome Rech}
\affiliation{Aix Marseille Universit\'e, CNRS, CPT, UMR 7332, 13288 Marseille, France}
\affiliation{Universit\'e de Toulon, CNRS, CPT, UMR 7332, 83957 La Garde, France}
\author{Thibaut Jonckheere}
\affiliation{Aix Marseille Universit\'e, CNRS, CPT, UMR 7332, 13288 Marseille, France}
\affiliation{Universit\'e de Toulon, CNRS, CPT, UMR 7332, 83957 La Garde, France}
\author{Thierry Martin}
\affiliation{Aix Marseille Universit\'e, CNRS, CPT, UMR 7332, 13288 Marseille, France}
\affiliation{Universit\'e de Toulon, CNRS, CPT, UMR 7332, 83957 La Garde, France}

\begin{abstract}
	We consider an electronic analog of the Hong-Ou-Mandel (HOM) interferometer,
	where two single electrons travel along opposite chiral edge states and collide at a quantum point contact.
	Studying the current noise, we show that  
	because of interactions between copropagating edge states, the degree of indistinguishability between
	the two electron wave packets is dramatically reduced, leading to reduced contrast
	for the HOM signal. This decoherence phenomenon strongly depends on the energy resolution of the packets. 
	Insofar as interactions cause charge fractionalization,
	we show that charge and neutral modes interfere with each other, leading to satellite dips or peaks in the current noise.
    Our calculations explain recent experimental results [E. Bocquillon, et al., Science {\bf 339}, 1054 (2013)] where an electronic HOM signal with reduced contrast
    was observed.
\end{abstract}

\maketitle

Electron quantum optics aims at transposing
quantum optics experiments -- such as Hanbury Brown-Twiss~\cite{HBT1,HBT2} (HBT) or
Hong-Ou-Mandel~\cite{HOM, Beugnon_nature_2006} (HOM) setups -- for the manipulation and measurement 
of single electrons propagating in quantum channels. 
Electrons differ from photons because of their 
statistics, the presence of the Fermi sea (FS) in
condensed matter systems and the fact that they interact. 
If controlling a single photon at a time was mastered long ago~\cite{1st_single_photon_source},
the emission of single electrons has only been achieved recently~\cite{SES_on_demand_ses, SES_current_correlations, energy_selective_excitations, Giblin_nat_comm_2012, Hermelin_nature_2011, McNeil_nature_2011}. 
This now allows us to implement in the integer quantum Hall effect (QHE) regime
the electronic analog of the HOM experiment~\cite{SES_HOM}, which, in optics, 
 measures the degree of indistinguishability between two photons
colliding on a beam splitter. Here, two electrons propagate along opposite edge states and collide at the location of a quantum point contact (QPC).   
Several theoretical works~\cite{BurkardLoss, Giovannetti_electronic_HOM, MoskaletsButtiker,two_particle_collider, HOM_cpt}
have addressed the outcome of this experiment 
at the single electron level (taking full account
of the statistics): the modulus of the current correlations
at the output of the QHE bar exhibits
a dip as a function of the time delay $\delta T$ between injections.
When $\delta T =0$, this dip extends down to 0. 
When the time delay is large enough, the two electrons no longer interfere and 
scatter independently at the QPC, and the current correlations correspond to the sum of the two HBT signals.

The puzzle with the recent experiment performed at a filling factor $\nu>1$, 
is that the HOM dip does not vanish as predicted for $\nu=1$. Here, we provide a theoretical framework for the experiment  
and we show that the interaction between quantum channels is responsible
for the observed effect. Indeed, at $\nu>1$, interactions dramatically change the nature of excitations,
leading to energy exchange between the channels and to charge fractionalization~\cite{MZI_nu2, Neder_fractionalization, Berg_fractional_charges, shot_noise_thermometry, KovrizhinChalker, shot_noise_nu2, Altimiras_spectroscopy, leSueur_energy_relaxation, EC_equilibration,degiovanni_plasmon_2010}.
Here, we consider a quantum Hall bar at $\nu=2$,
in the strong coupling regime and at finite temperature
($\Theta \sim 100~\mathrm{mK}$, following the experiment). 
The effect of interactions between edge states is probed by comparison with previous
results obtained at $\nu=1$ without interactions~\cite{HOM_cpt}.\\

\paragraph{Charge fractionalization.--}

\begin{figure}
	\includegraphics[scale=.85]{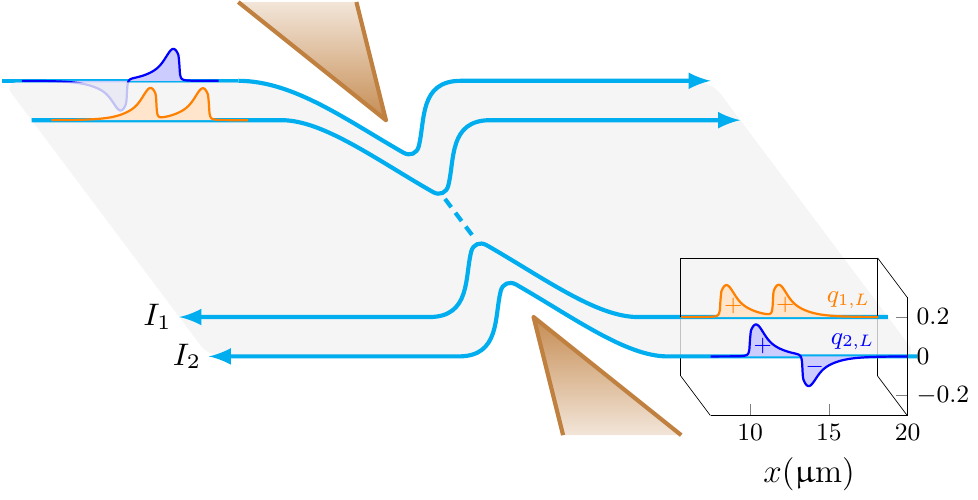}
	\caption{\label{setup} The setups: two opposite edge states, each made out of two interacting copropagating channels, meet at a QPC. 
An electronic wave packet is injected on both incoming outer channels. 
(Main) Setup 1: backscattering occurs  for outer channels. 
 (Left inset) Setup 2: backscattering occurs for inner channels.
(Right inset) Electron density as a function of position for an energy resolved packet 
 imaged after propagating on a $5~\mathrm{\mu m}$ length, revealing the presence of two modes composed each of two $\oplus/\ominus$ excitations.
	 }
\end{figure}

On each edge, the two copropagating channels are coupled
via Coulomb interaction modeled as a short-range interaction $H_\text{int}$.
Label $j=1,2$ identifies outer and inner channels while $r=R,L$ stands for right and left
moving ones. The electronic annihilation operator at position $x$ and
time $t$ reads $\psi_{j,r} (x,t) = U_r/\sqrt{2 \pi a}\: e^{i \phi_{j,r} (x,t)}$
with $\phi$ the chiral Luttinger bosonic field~\cite{bosonization_refermionization}, $U_r$ a Klein factor
and $a$ a cutoff parameter.
The Hamiltonian is the sum of its kinetic and interaction contributions:
\begin{align}
	H_\text{kin} &= \sum_{j=1,2} v_j \frac{\hbar}{\pi} \sum_{r=R,L} \int d x (\partial_x \phi_{j,r} )^2\\
	H_\text{int} &= 2 u \frac{\hbar}{\pi} \sum_r \int d x (\partial_x \phi_{1,r}) (\partial_x \phi_{2,r}) \label{Hint}
\end{align}
where $u$ describes the interaction strength.
The charge density operator is
$q_{j,r} (x,t)= e/\pi \partial_x \phi_{j,r} (x,t)$, and thus, Eq. (\ref{Hint}) describes
a local capacitive coupling between  copropagating channels. Intrachannel interactions can
be taken into account by a renormalization of the velocities $v_{1,2}$.
The full interacting problem can now be diagonalized with a rotation of angle
$\theta$ defined as $\tan (2 \theta) = 2 u /(v_1-v_2)$ which expresses 
the coupling strength. Since the strong
coupling regime $\theta=\pi/4$ seems to be the most relevant
experimentally~\cite{Bocquillon_interedge},
we focus on $v_1=v_2=v_F=1$.
The Hamiltonian is next expressed in terms
of the rotated fields
$\phi_\pm =( \phi_2 \pm \phi_1)/\sqrt{2}$ as
$H=\hbar/\pi \sum_r \int d x\: v_+ (\partial_x \phi_{+,r} )^2 + v_- (\partial_x \phi_{-,r} )^2$.
It describes the free propagation of the two collective modes: a fast charge mode traveling with velocity $v_+=1 + u$ and a slow neutral mode propagating at $v_-=1 - u$. Along a given edge, each of these modes can be viewed as two separate excitations propagating on the different channels composing the edge, and characterized by the charge they carry ($\oplus$ or $\ominus$). 
The single electron source is modeled through the injection of single electronic wave packets along the edges at a given distance from the QPC, which amounts to calculating all average values over a prepared state~\cite{HOM_cpt}. In order to be as close as can be to the experiment~\cite{SES_HOM},
electrons are injected along the outer channel as exponential wave packets in real space
$\varphi_L (x) = \sqrt{ 2 \Gamma} e^{-i \epsilon_0 x} e^{-\Gamma x} \theta(x)$.
The higher the energy resolution $\gamma=\epsilon_0/\Gamma$ of the wave packet, the more it decoheres, as its energy distribution is more distorted from its original shape.
Propagation in real space simply amounts to translating the charge and neutral modes with their respective velocity even though their energy profile is strongly modified along the edge.

\paragraph{HBT.--}
We partition at a QPC
the excitations propagating along the left-moving edge following the injection along the outer channel of an exponential wave packet at position $x=L$.
 Depending on which counter-propagating channels are connected at the QPC, one can distinguish two setups (see Fig.~\ref{setup}) which we hereby label as setup 1 and 2, corresponding, respectively, to the partitioning of the outer ($s=1$) or the inner channel ($s=2$). Both these setups can be realized in practice and although experimental investigations have focused so far on setup 1, we will be considering them both.

The quantity of interest is the zero-frequency current correlations~\cite{Martin_Houches, Martin_1992} measured on the partitioned channel:
$S_\text{HBT}=\int d t d t' \langle I_s (t) I_s (t') \rangle - \langle I_s (t) \rangle \langle I_s (t') \rangle$,
where the averages are performed on the prepared state
$|\phi_L \rangle = \int d x \varphi_L (x) \psi^\dagger_{1,L} (x+L,0) |0\rangle$.
All integrals are computed
from $-\infty$ to $+\infty$, and
the FS contribution has been removed.
The linear dispersion of the edges
allows us to compute
the noise at the immediate output of the QPC, without loss of generality.
The latter is described by its scattering matrix which relates
the outgoing fields along the partitioned channel to the incoming ones
\begin{equation}
	\begin{pmatrix}	\psi_{s,R} \\ \psi_{s,L} \end{pmatrix}^{\mathrm{outgoing}}
	= \begin{pmatrix}
		\sqrt{{\cal T}} & i \sqrt{{\cal R}}\\
		i \sqrt{{\cal R}} & \sqrt{{\cal T }}
	\end{pmatrix} \begin{pmatrix}
		\psi_{s,R} \\
		\psi_{s,L}
	\end{pmatrix}^{\mathrm{incoming}} \label{scattering}
\end{equation}
where ${\cal T}$ and ${\cal R}=1-{\cal T}$ are the transmission and reflection
probabilities. 
This scattering approach, albeit used
for interacting fermion fields,
is justified because both the interaction and the tunneling are purely local. The QPC is, thus, not included in the interaction region, and fermions
are locally free at this location. The proof, based on refermionization,
is given in the Supplemental Material (SM)~\cite{SM}.
Equation~\eqref{scattering} allows us to express the noise in terms
of the incoming operators only~\cite{SES_tomography}. Since the injection
process is noiseless within our model, we are left with
\begin{align}
	S_\text{HBT} = - e^2 {\cal R T} \int d t d t' \langle \psi_{s,R}^\dagger (t) \psi_{s,R} (t') \rangle \langle \psi_{s,L} (t)
	\psi_{s,L}^\dagger (t') \rangle \nonumber\\
	+\langle \psi_{s,L}^\dagger (t) \psi_{s,L} (t') \rangle \langle
	\psi_{s,R} (t) \psi_{s,R}^\dagger (t') \rangle
\end{align}
where all quantities are computed at the input of the QPC.
The averages are expressed in terms of the fast and slow Green's functions
of the bosonic fields
\begin{widetext}
	\begin{equation}
		S_\text{HBT} = -  \frac{2 e^2 {\cal R T}}{(2 \pi a)^3 {\cal N}} \mathrm{Re} \left\{ 
 \int \!\!  d y_L  d z_L  \; \varphi_L (y_L) \varphi_L^* (z_L) g(0,z_L-y_L) 
   \int \!\! dt \, d \tau \, \mathrm{Re} \left[ g(\tau,0)^2 \right] \left[ \frac{h_s (t;y_L+L,z_L+L)}{h_s (t+\tau;y_L+L,z_L+L)} -1 \right] \right\} ,
\label{HBT_noise}
	\end{equation}
\end{widetext}
where ${\cal N}= \langle \phi_L | \phi_L \rangle$, $s=1,2$ is the setup considered and 
\begin{align*}
g(t,x)&= \left[ \frac{\sinh \left( i \frac{\pi a}{\beta v_+} \right)}{\sinh \left(  \frac{ia + v_+ t - x}{\beta v_+ /\pi} \right)}  \frac{\sinh \left( i \frac{\pi a}{\beta v_-} \right)}{\sinh \left( \frac{ia + v_- t - x}{\beta v_- /\pi} \right)} \right]^{1/2} , \\
h_s (t;x,y) &= \left[ \frac{\sinh \left( \frac{ia - v_+ t + x}{\beta v_+ / \pi} \right)}{\sinh \left(\frac{ia + v_+ t - y}{\beta v_+/\pi} \right)} \right]^{\frac{1}{2}} \left[ \frac{\sinh \left( \frac{ia - v_- t + x}{\beta v_- / \pi} \right)}{\sinh \left( \frac{ia + v_- t - y}{\beta v_- /\pi} \right)} \right]^{s-\frac{3}{2}} .
\end{align*}
Numerical integration is handled with a quasi-Monte Carlo algorithm using importance sampling~\cite{Cuba} (details are given in the SM~\cite{SM}).
Since the partition noise 
counts the number of particle-hole excitations~\cite{two_particle_collider}, the absolute value of the noise
increases with the interaction strength and the  energy resolution of the packet:
 as a single electron injected above the FS relaxes,
it creates particle-hole pairs near the Fermi energy~\cite{Degiovanni_decoherence} which scatter at the QPC.
These spurious excitations are more numerous for an energy-resolved packet, resulting in a larger noise.
The dependence of the noise on $L$ is then governed by two opposing effects.
While the eigenmodes are dragged apart when $L$ rises, the number of particle-hole pairs
increases and finally diverges as $\log L$ at zero temperature~\cite{Lee_orthogonality_catastrophe} (see SM),
leading to the same divergence in
the noise.
However, at finite temperature, $S_\text{HBT}$ is dramatically reduced because
of antibunching with thermal excitations at the output of the QPC~\cite{SES_HBT}. 
This tends to minimize the contribution to the noise from low-energy quasiparticles, which were dominant at $\Theta=0$.
The finite temperature, thus, acts as a low-energy cutoff, washing out the length dependence of the noise, which typically is constant for $L\ge2~\mathrm{\mu m}$, at $\Theta \sim 100~\mathrm{mK}$.

\paragraph{HOM.--}
The next step is to make two wave packets collide at the QPC.
The prepared state is now  $|\phi_R \rangle \otimes |\phi_L \rangle$,
where two electrons are injected on the outer channel of the counter-propagating arms.
For  simplicity, we consider injections at symmetric positions $\pm L$, focusing on the interference between identical wave packets, $\varphi_R (x)=\varphi_L (-x)$. 
The expression for the noise, Eq.~(\ref{HBT_noise}), is modified as
\begin{widetext}
	\begin{align}
		S_\text{HOM} (\delta T) = - \frac{2e ^2 {\cal R T}}{(2 \pi a)^4 {\cal N}}
		     \mathrm{Re} & \left\{ \int \!\! d y_L d z_L \; \varphi_L(y_L) \varphi_L^*(z_L)  g (0,z_L-y_L) 
		                           \int \!\! d y_R d z_R \; \varphi_R(y_R) \varphi_R^*(z_R)  g(0,y_R-z_R) \right. \nonumber\\
		&\left. \times   \int \!\! d \tau \, \mathrm{Re} \left[ g(\tau,0)^2 \right]  \int \!\! dt \left[ \frac{h_s(t;y_L+L,z_L+L)}{h_s (t+\tau;y_L+L,z_L+L)} \frac{h_s (t+\tau-\delta T;L-y_R,L-z_R)}{h_s (t-\delta T;L-y_R,L-z_R)}   - 1 \right] \right\} .
		 \label{HOM_noise}
	\end{align}
\end{widetext}

\begin{figure*}
	\begin{ruledtabular}
		\begin{tabular}{r r}
			\raisebox{.09\height}{\includegraphics[scale=0.7]{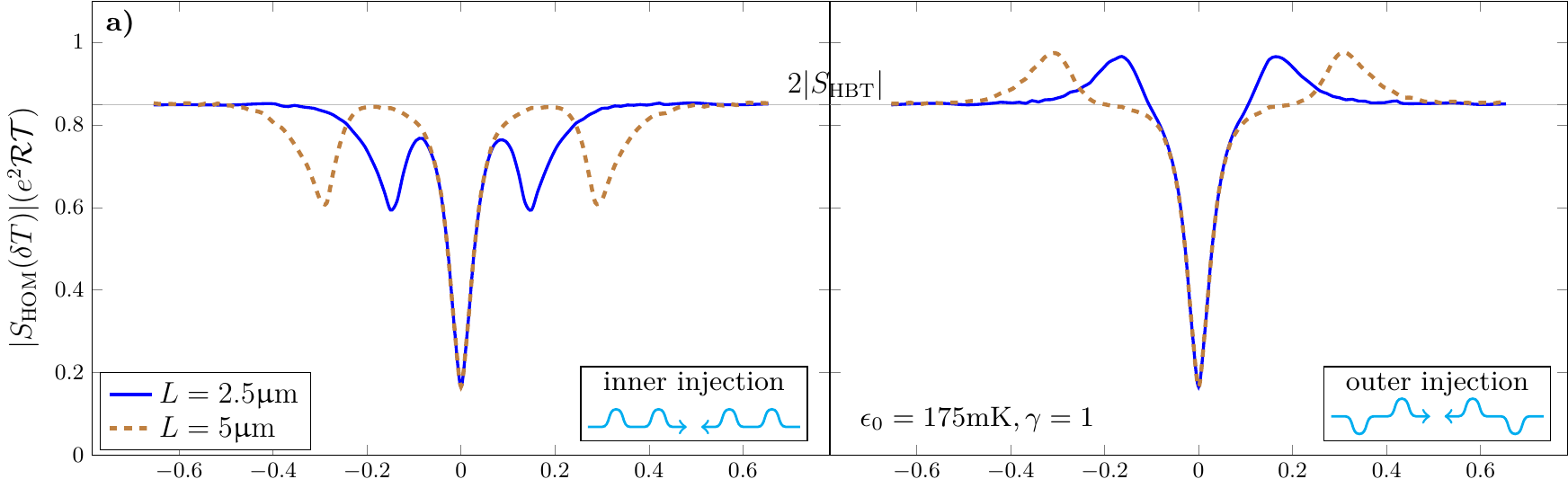}}&
			\includegraphics[scale=0.7]{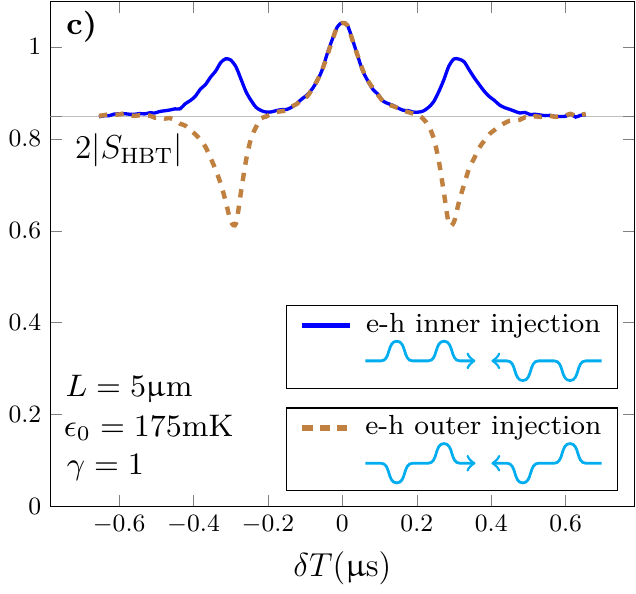}
			\\
			\includegraphics[scale=0.7]{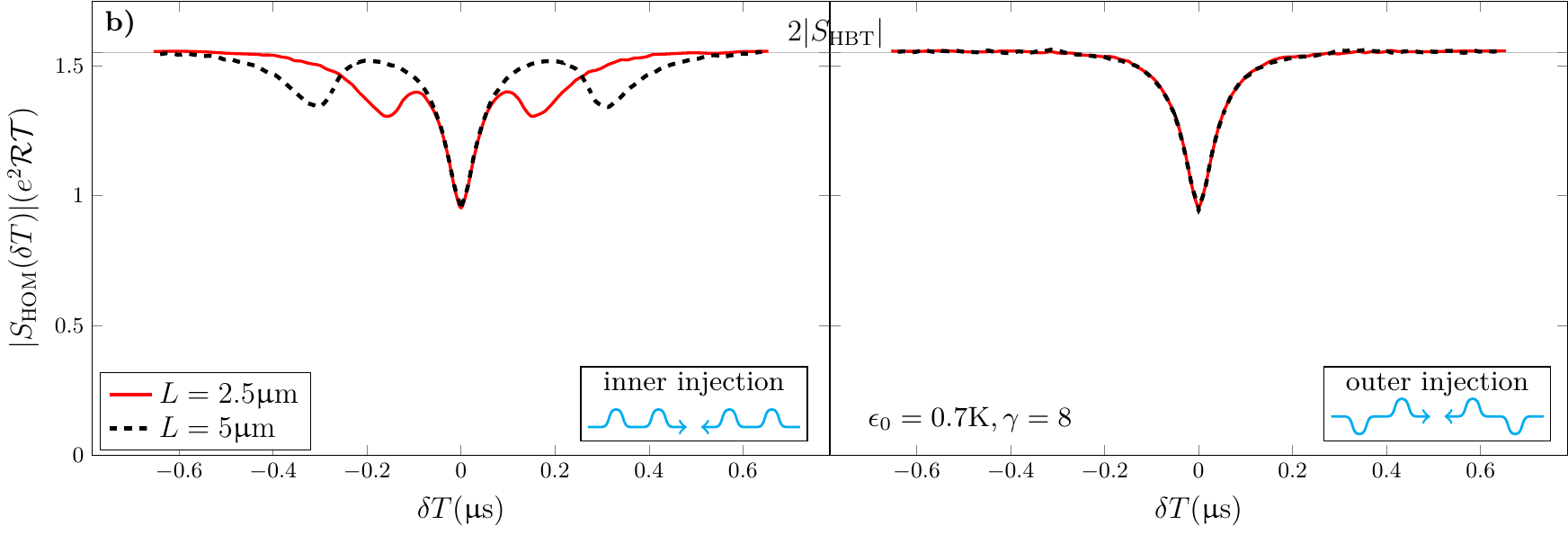}&
			\includegraphics[scale=0.7]{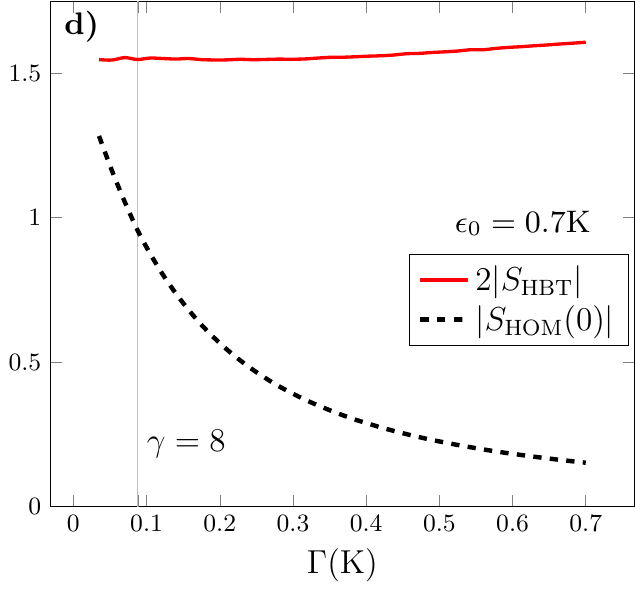}
		\end{tabular}
	\end{ruledtabular}
	\caption{\label{HOM} (a)-(c)Modulus of $S_\text{HOM}$ in units of
	$e^2 {\cal R T}$
	as a function of the time delay $\delta T$,
	for setups 1 and 2.
	(a)Packets wide in energy ($\gamma=1$, $\epsilon_0=175~\mathrm{mK}$ is the injection energy).
	(b)Energy-resolved packets ($\gamma=8$)
	give a lower contrast. Results for setup 1 reveal a triple dip structure, while for setup 2 we obtain a peak-dip-peak structure,
	with vanishingly small peaks in the case of energy-resolved packets.
	(c)Electron-hole interference: an electron has been injected
	on the right moving arm and a hole on the left moving one.
	(d)Modulus of $S_\text{HOM} (0)$ and $2 S_\text{HBT}$ in units of
	$e^2 {\cal R T}$ as a function of $\Gamma$, for $\epsilon_0=0.7\kelvin$.
	In all plots, $u=0.5$ and $\Theta=0.1 \kelvin$.
	}
\end{figure*}

As the time delay $\delta T$ between the right- and left-moving electron is varied,
we find three characteristic signatures in the noise (see Figs.~\ref{HOM}(a)--\ref{HOM}(b)).
At $\delta T = 0$, a central dip appears, with a depth which depends strongly on the injected packet energy resolution, but very little on the actual setup considered.
At $\delta T= \pm 2 L u /(1-u^2)$, side structures emerge symmetrically with respect to the central dip, with a depth and shape that is again conditioned by the energy resolution of the wave packet. Interestingly, these results vary critically between setups, as these side structures manifest as dips for setup 1, but peaks for setup 2.
Away from these three features, $S_{\text{HOM}}$ saturates at twice the HBT noise:  electrons injected on the two incoming arms scatter independently at the QPC.

This interference pattern is interpreted in terms of the different excitations propagating along the partitioned edge channel.  After injection, the electron fractionalizes into two modes. The fast charge mode is composed of two $\oplus$ excitations. The slow neutral mode is made out of a $\oplus$ excitation propagating along the injection channel and a $\ominus$ excitation traveling along the copropagating channel.
The central dip, which corresponds to the symmetric situation of synchronized injections, thus probes the interference of excitations with the same velocity and charge: two fast $\oplus$ excitations then two slow $\oplus$ or $\ominus$ excitations (depending on the setup). These identical excitations interfere destructively, leading to a reduction of the noise (in absolute value), thus, producing a dip.
Note that the bottom of this dip is practically insensitive to the chosen setup (and thus, to the partitioned channel) signaling that the interference between identical excitations is independent on the charge they carry.

A striking difference with the $\nu=1$ case is that the central dip never reaches down to $0$, as  observed experimentally~\cite{SES_HOM}.
The depth of this dip is actually a probing tool of the degree of indistinguishability
between the colliding excitations~\cite{Feve_flying_qubits}. Our present work suggests that because of the strong interchannel coupling,
some coherence is lost in the copropagating channels, and the Coulomb-induced decoherence leads to this characteristic loss of contrast for the HOM dip.  
This effect gets more pronounced for further energy-resolved packets.
As depicted in Figs.~\ref{HOM}(a)--\ref{HOM}(b), while for ``wide'' packets in energy ($\gamma=1$) the contrast is still pretty good, $\eta \sim 0.8$, the loss of contrast can be dramatic for energy-resolved packets, with $\eta \sim 0.4$ for $\gamma=8$. The same trend is observed in experiments~\cite{these_Bocquillon}. Results for the dip depth at $\Theta=0$ are discussed in the SM~\cite{SM}.

Adjusting $\delta T$ appropriately, one can also probe interferences between excitations that have different velocities. This effect is responsible for the side structures appearing in the noise: at $\delta T=2 L u/(1-u^2)$, the fast right-moving excitation and the slow left-moving one reach the QPC at the same time~\footnote{And similarly, the feature at $\delta T=-2 L u/(1-u^2)$ corresponds to the collision between a slow right-moving excitation and a fast left-moving one.}.
In setup 1, these lateral structures correspond to the collision of two $\oplus$ excitations, which interfere destructively, as argued earlier, leading to dips. Their depth is, however, less than half the one of the central dip. This can be attributed to the velocity mismatch between interfering excitations, as it indicates that they are more distinguishable. 
More interestingly, setup 2 allows us to probe the encounter of excitations with opposite charge (and different velocities), which is expected to lead to constructive interference. This is consistent with the occurrence of lateral peaks in our calculations [see Fig.~\ref{HOM}(a), right column]. It is reminiscent of what was predicted in the $\nu=1$ case for electron-hole HOM interferometry~\cite{HOM_cpt}. The side peaks are more pronounced for small $\gamma$, and become vanishingly small for larger values of the energy resolution, signaling a nontrivial dependence on the packet energy content.

All these lateral dips and peaks are asymmetric as a consequence of the velocity difference
between excitations. Typically, the slope is steeper for smaller $| \delta T |$. 
This asymmetry is similar to the one encountered in the noninteracting $\nu=1$ case for interfering packets with different shapes, where a broad right-moving packet in space collides onto a thin left-moving one~\cite{HOM_cpt}.

Our approach is general enough to be extended to regimes that have yet to be explored experimentally, such as  electron-hole interferometry, where an electron is injected on one edge, while a hole is injected on the other edge
[see Fig.~\ref{HOM}(c)]. 
There, we recover three structures in the noise.
For both setups, at $\delta T=0$ $\oplus$ excitations interfere constructively with $\ominus$ excitations, leading to a central peak. 
However, while setup 1 shows lateral peaks produced by interfering oppositely charged excitations, setup 2 probes the interference of excitations carrying the same charge, leading to lateral dips. Results concerning the dependence on the energy width of the packet
are presented in Fig.~\ref{HOM}(d), for an injection well above
the FS ($\epsilon_0=0.7\kelvin$). First, the HBT noise does not depend much
on $\Gamma$ but the central dip sinks drastically as $\Gamma$ is increased,
leading to a much higher contrast. The more resolved a packet is in energy,
the more it decoheres and the worse the contrast.


To conclude, strong coupling between the copropagating
channels accounts for a sensible loss of contrast of the HOM central dip
as observed in the experiment~\cite{SES_HOM}.
This reduction factor strongly depends on the energy resolution of
the emitted packets and is directly related to decoherence. Moreover, fast and slow modes do interfere with each other and, depending on the charge carried by the colliding excitations, produce smaller asymmetric side dips or side peaks.
While these have not yet been observed, upcoming experiments with better resolution should reveal such signatures, especially when operating at lower excitation frequency, thus, accessing a wider interval of $\delta T$.
This constitutes an important test, along with the expected variations in $L$ and $\Gamma$.
The predicted behavior as $L$ is varied could be
checked if lateral gates were added to the setup, modifying 
the propagation path before the QPC.
Measurements with different injection energies and packet widths are already being processed.
Extensions to long-range interactions~\cite{Bocquillon_interedge} and to the fractional QHE are considered.

\begin{acknowledgments}
We acknowledge the support of
ANR-2010-BLANC-0412, ANR-11-LABX-0033, ANR-11-IDEX-0001-02
 and the M\'esocentre d'Aix-Marseille Universit\'e.
We are grateful to P. Degiovanni, G. F\`eve, and F. Portier for useful discussions.

\end{acknowledgments}

During the process of review of the manuscript, we came aware of the work of~\cite{Dubois_2013}, which demonstrates a new kind of electron source relevant for the present work.

\bibliography{biblio4}

\begin{thebibliography}{40}
\expandafter\ifx\csname natexlab\endcsname\relax\def\natexlab#1{#1}\fi
\expandafter\ifx\csname bibnamefont\endcsname\relax
  \def\bibnamefont#1{#1}\fi
\expandafter\ifx\csname bibfnamefont\endcsname\relax
  \def\bibfnamefont#1{#1}\fi
\expandafter\ifx\csname citenamefont\endcsname\relax
  \def\citenamefont#1{#1}\fi
\expandafter\ifx\csname url\endcsname\relax
  \def\url#1{\texttt{#1}}\fi
\expandafter\ifx\csname urlprefix\endcsname\relax\def\urlprefix{URL }\fi
\providecommand{\bibinfo}[2]{#2}
\providecommand{\eprint}[2][]{\url{#2}}

\bibitem[{\citenamefont{Brown and Twiss}(1957)}]{HBT1}
\bibinfo{author}{\bibfnamefont{R.~H.} \bibnamefont{Brown}} \bibnamefont{and}
  \bibinfo{author}{\bibfnamefont{R.~Q.} \bibnamefont{Twiss}},
  \bibinfo{journal}{Proc. R. Soc. A} \textbf{\bibinfo{volume}{242}},
  \bibinfo{pages}{300} (\bibinfo{year}{1957}).

\bibitem[{\citenamefont{Brown and Twiss}(1958)}]{HBT2}
\bibinfo{author}{\bibfnamefont{R.~H.} \bibnamefont{Brown}} \bibnamefont{and}
  \bibinfo{author}{\bibfnamefont{R.~Q.} \bibnamefont{Twiss}},
  \bibinfo{journal}{Proc. R. Soc. A} \textbf{\bibinfo{volume}{243}},
  \bibinfo{pages}{291} (\bibinfo{year}{1958}).

\bibitem[{\citenamefont{Hong et~al.}(1987)\citenamefont{Hong, Ou, and
  Mandel}}]{HOM}
\bibinfo{author}{\bibfnamefont{C.~K.} \bibnamefont{Hong}},
  \bibinfo{author}{\bibfnamefont{Z.~Y.} \bibnamefont{Ou}}, \bibnamefont{and}
  \bibinfo{author}{\bibfnamefont{L.}~\bibnamefont{Mandel}},
  \bibinfo{journal}{Phys. Rev. Lett.} \textbf{\bibinfo{volume}{59}},
  \bibinfo{pages}{2044} (\bibinfo{year}{1987}).

\bibitem[{\citenamefont{Beugnon et~al.}(2006)\citenamefont{Beugnon, Jones,
  Dingjan, Darquié, Messin, Browaeys, and Grangier}}]{Beugnon_nature_2006}
\bibinfo{author}{\bibfnamefont{J.}~\bibnamefont{Beugnon}},
  \bibinfo{author}{\bibfnamefont{M.~P.~A.} \bibnamefont{Jones}},
  \bibinfo{author}{\bibfnamefont{J.}~\bibnamefont{Dingjan}},
  \bibinfo{author}{\bibfnamefont{B.}~\bibnamefont{Darquié}},
  \bibinfo{author}{\bibfnamefont{G.}~\bibnamefont{Messin}},
  \bibinfo{author}{\bibfnamefont{A.}~\bibnamefont{Browaeys}}, \bibnamefont{and}
  \bibinfo{author}{\bibfnamefont{P.}~\bibnamefont{Grangier}},
  \bibinfo{journal}{Nature (London)} \textbf{\bibinfo{volume}{440}},
  \bibinfo{pages}{779} (\bibinfo{year}{2006}).

\bibitem[{\citenamefont{Clauser}(1974)}]{1st_single_photon_source}
\bibinfo{author}{\bibfnamefont{J.~F.} \bibnamefont{Clauser}},
  \bibinfo{journal}{Phys. Rev. D} \textbf{\bibinfo{volume}{9}},
  \bibinfo{pages}{853} (\bibinfo{year}{1974}).

\bibitem[{\citenamefont{F{\`e}ve et~al.}(2007)\citenamefont{F{\`e}ve, Mah{\'e},
  Berroir, Kontos, Pla\c{c}ais, Glatti, Cavanna, Etienne, and
  Jin}}]{SES_on_demand_ses}
\bibinfo{author}{\bibfnamefont{G.}~\bibnamefont{F{\`e}ve}},
  \bibinfo{author}{\bibfnamefont{A.}~\bibnamefont{Mah{\'e}}},
  \bibinfo{author}{\bibfnamefont{J.-M.} \bibnamefont{Berroir}},
  \bibinfo{author}{\bibfnamefont{T.}~\bibnamefont{Kontos}},
  \bibinfo{author}{\bibfnamefont{B.}~\bibnamefont{Pla\c{c}ais}},
  \bibinfo{author}{\bibfnamefont{D.~C.} \bibnamefont{Glatti}},
  \bibinfo{author}{\bibfnamefont{A.}~\bibnamefont{Cavanna}},
  \bibinfo{author}{\bibfnamefont{B.}~\bibnamefont{Etienne}}, \bibnamefont{and}
  \bibinfo{author}{\bibfnamefont{Y.}~\bibnamefont{Jin}},
  \bibinfo{journal}{Science} \textbf{\bibinfo{volume}{316}},
  \bibinfo{pages}{196404} (\bibinfo{year}{2007}).

\bibitem[{\citenamefont{Mah{\'e} et~al.}(2010)\citenamefont{Mah{\'e},
  Parmentier, Bocquillon, Berroir, Glattli, Kontos, Pla\c{c}ais, F{\`e}ve,
  Cavanna, and Jin}}]{SES_current_correlations}
\bibinfo{author}{\bibfnamefont{A.}~\bibnamefont{Mah{\'e}}},
  \bibinfo{author}{\bibfnamefont{F.~D.} \bibnamefont{Parmentier}},
  \bibinfo{author}{\bibfnamefont{E.}~\bibnamefont{Bocquillon}},
  \bibinfo{author}{\bibfnamefont{J.-M.} \bibnamefont{Berroir}},
  \bibinfo{author}{\bibfnamefont{D.~C.} \bibnamefont{Glattli}},
  \bibinfo{author}{\bibfnamefont{T.}~\bibnamefont{Kontos}},
  \bibinfo{author}{\bibfnamefont{B.}~\bibnamefont{Pla\c{c}ais}},
  \bibinfo{author}{\bibfnamefont{G.}~\bibnamefont{F{\`e}ve}},
  \bibinfo{author}{\bibfnamefont{A.}~\bibnamefont{Cavanna}}, \bibnamefont{and}
  \bibinfo{author}{\bibfnamefont{Y.}~\bibnamefont{Jin}},
  \bibinfo{journal}{Phys. Rev. B} \textbf{\bibinfo{volume}{82}},
  \bibinfo{pages}{201309} (\bibinfo{year}{2010}).

\bibitem[{\citenamefont{Leicht et~al.}(2011)\citenamefont{Leicht, Mirovsky,
  Kaestner, Hohls, Kashcheyevs, Kurganova, Zeitler, Weimann, Pierz, and
  Schumacher}}]{energy_selective_excitations}
\bibinfo{author}{\bibfnamefont{C.}~\bibnamefont{Leicht}},
  \bibinfo{author}{\bibfnamefont{P.}~\bibnamefont{Mirovsky}},
  \bibinfo{author}{\bibfnamefont{B.}~\bibnamefont{Kaestner}},
  \bibinfo{author}{\bibfnamefont{F.}~\bibnamefont{Hohls}},
  \bibinfo{author}{\bibfnamefont{V.}~\bibnamefont{Kashcheyevs}},
  \bibinfo{author}{\bibfnamefont{E.~V.} \bibnamefont{Kurganova}},
  \bibinfo{author}{\bibfnamefont{U.}~\bibnamefont{Zeitler}},
  \bibinfo{author}{\bibfnamefont{T.}~\bibnamefont{Weimann}},
  \bibinfo{author}{\bibfnamefont{K.}~\bibnamefont{Pierz}}, \bibnamefont{and}
  \bibinfo{author}{\bibfnamefont{H.~W.} \bibnamefont{Schumacher}},
  \bibinfo{journal}{Semicond. Sci. Technol.} \textbf{\bibinfo{volume}{26}},
  \bibinfo{pages}{055010} (\bibinfo{year}{2011}).

\bibitem[{\citenamefont{Giblin et~al.}(2012)\citenamefont{Giblin, Kataoka,
  Fletcher, See, Janssen, Griffiths, Jones, Farrer, and
  Ritchie}}]{Giblin_nat_comm_2012}
\bibinfo{author}{\bibfnamefont{S.}~\bibnamefont{Giblin}},
  \bibinfo{author}{\bibfnamefont{M.}~\bibnamefont{Kataoka}},
  \bibinfo{author}{\bibfnamefont{J.}~\bibnamefont{Fletcher}},
  \bibinfo{author}{\bibfnamefont{P.}~\bibnamefont{See}},
  \bibinfo{author}{\bibfnamefont{T.}~\bibnamefont{Janssen}},
  \bibinfo{author}{\bibfnamefont{J.}~\bibnamefont{Griffiths}},
  \bibinfo{author}{\bibfnamefont{G.}~\bibnamefont{Jones}},
  \bibinfo{author}{\bibfnamefont{I.}~\bibnamefont{Farrer}}, \bibnamefont{and}
  \bibinfo{author}{\bibfnamefont{D.}~\bibnamefont{Ritchie}},
  \bibinfo{journal}{Nat. Commun.} \textbf{\bibinfo{volume}{3}},
  \bibinfo{pages}{930} (\bibinfo{year}{2012}).

\bibitem[{\citenamefont{Hermelin et~al.}(2011)\citenamefont{Hermelin, Takada,
  Yamamoto, Tarucha, Wieck, Saminadayar, Bäuerle, and
  Meunier}}]{Hermelin_nature_2011}
\bibinfo{author}{\bibfnamefont{S.}~\bibnamefont{Hermelin}},
  \bibinfo{author}{\bibfnamefont{S.}~\bibnamefont{Takada}},
  \bibinfo{author}{\bibfnamefont{M.}~\bibnamefont{Yamamoto}},
  \bibinfo{author}{\bibfnamefont{S.}~\bibnamefont{Tarucha}},
  \bibinfo{author}{\bibfnamefont{A.~D.} \bibnamefont{Wieck}},
  \bibinfo{author}{\bibfnamefont{L.}~\bibnamefont{Saminadayar}},
  \bibinfo{author}{\bibfnamefont{C.}~\bibnamefont{Bäuerle}}, \bibnamefont{and}
  \bibinfo{author}{\bibfnamefont{T.}~\bibnamefont{Meunier}},
  \bibinfo{journal}{Nature (London)} \textbf{\bibinfo{volume}{477}},
  \bibinfo{pages}{435} (\bibinfo{year}{2011}).

\bibitem[{\citenamefont{McNeil et~al.}(2011)\citenamefont{McNeil, Kataoka,
  Ford, Barnes, Anderson, Jones, Farrer, and Ritchie}}]{McNeil_nature_2011}
\bibinfo{author}{\bibfnamefont{R.~P.~G.} \bibnamefont{McNeil}},
  \bibinfo{author}{\bibfnamefont{M.}~\bibnamefont{Kataoka}},
  \bibinfo{author}{\bibfnamefont{C.~J.~B.} \bibnamefont{Ford}},
  \bibinfo{author}{\bibfnamefont{C.~H.~W.} \bibnamefont{Barnes}},
  \bibinfo{author}{\bibfnamefont{D.}~\bibnamefont{Anderson}},
  \bibinfo{author}{\bibfnamefont{G.~A.~C.} \bibnamefont{Jones}},
  \bibinfo{author}{\bibfnamefont{I.}~\bibnamefont{Farrer}}, \bibnamefont{and}
  \bibinfo{author}{\bibfnamefont{D.~A.} \bibnamefont{Ritchie}},
  \bibinfo{journal}{Nature (London)} \textbf{\bibinfo{volume}{477}},
  \bibinfo{pages}{439} (\bibinfo{year}{2011}).

\bibitem[{\citenamefont{Bocquillon
  et~al.}(2013{\natexlab{a}})\citenamefont{Bocquillon, Freulon, Berroir,
  Degiovanni, Pla\c{c}ais, Cavanna, Jin, and F{\`e}ve}}]{SES_HOM}
\bibinfo{author}{\bibfnamefont{E.}~\bibnamefont{Bocquillon}},
  \bibinfo{author}{\bibfnamefont{V.}~\bibnamefont{Freulon}},
  \bibinfo{author}{\bibfnamefont{J.-M.} \bibnamefont{Berroir}},
  \bibinfo{author}{\bibfnamefont{P.}~\bibnamefont{Degiovanni}},
  \bibinfo{author}{\bibfnamefont{B.}~\bibnamefont{Pla\c{c}ais}},
  \bibinfo{author}{\bibfnamefont{A.}~\bibnamefont{Cavanna}},
  \bibinfo{author}{\bibfnamefont{Y.}~\bibnamefont{Jin}}, \bibnamefont{and}
  \bibinfo{author}{\bibfnamefont{G.}~\bibnamefont{F{\`e}ve}},
  \bibinfo{journal}{Science} \textbf{\bibinfo{volume}{339}},
  \bibinfo{pages}{1054} (\bibinfo{year}{2013}{\natexlab{a}}).

\bibitem[{\citenamefont{Burkard and Loss}(2003)}]{BurkardLoss}
\bibinfo{author}{\bibfnamefont{G.}~\bibnamefont{Burkard}} \bibnamefont{and}
  \bibinfo{author}{\bibfnamefont{D.}~\bibnamefont{Loss}},
  \bibinfo{journal}{Phys. Rev. Lett.} \textbf{\bibinfo{volume}{91}},
  \bibinfo{pages}{087903} (\bibinfo{year}{2003}).

\bibitem[{\citenamefont{Giovannetti et~al.}(2006)\citenamefont{Giovannetti,
  Frustaglia, Taddei, and Fazio}}]{Giovannetti_electronic_HOM}
\bibinfo{author}{\bibfnamefont{V.}~\bibnamefont{Giovannetti}},
  \bibinfo{author}{\bibfnamefont{D.}~\bibnamefont{Frustaglia}},
  \bibinfo{author}{\bibfnamefont{F.}~\bibnamefont{Taddei}}, \bibnamefont{and}
  \bibinfo{author}{\bibfnamefont{R.}~\bibnamefont{Fazio}},
  \bibinfo{journal}{Phys. Rev. B} \textbf{\bibinfo{volume}{74}},
  \bibinfo{pages}{115315} (\bibinfo{year}{2006}).

\bibitem[{\citenamefont{Moskalets and B{\"u}ttiker}(2011)}]{MoskaletsButtiker}
\bibinfo{author}{\bibfnamefont{M.}~\bibnamefont{Moskalets}} \bibnamefont{and}
  \bibinfo{author}{\bibfnamefont{M.}~\bibnamefont{B{\"u}ttiker}},
  \bibinfo{journal}{Phys. Rev. B} \textbf{\bibinfo{volume}{83}},
  \bibinfo{pages}{035316} (\bibinfo{year}{2011}).

\bibitem[{\citenamefont{Ol'khovskaya et~al.}(2008)\citenamefont{Ol'khovskaya,
  Splettstoesser, Moskalets, and B{\"u}ttiker}}]{two_particle_collider}
\bibinfo{author}{\bibfnamefont{S.}~\bibnamefont{Ol'khovskaya}},
  \bibinfo{author}{\bibfnamefont{J.}~\bibnamefont{Splettstoesser}},
  \bibinfo{author}{\bibfnamefont{M.}~\bibnamefont{Moskalets}},
  \bibnamefont{and}
  \bibinfo{author}{\bibfnamefont{M.}~\bibnamefont{B{\"u}ttiker}},
  \bibinfo{journal}{Phys. Rev. Lett.} \textbf{\bibinfo{volume}{101}},
  \bibinfo{pages}{166802} (\bibinfo{year}{2008}).

\bibitem[{\citenamefont{Jonckheere et~al.}(2012)\citenamefont{Jonckheere, Rech,
  Wahl, and Martin}}]{HOM_cpt}
\bibinfo{author}{\bibfnamefont{T.}~\bibnamefont{Jonckheere}},
  \bibinfo{author}{\bibfnamefont{J.}~\bibnamefont{Rech}},
  \bibinfo{author}{\bibfnamefont{C.}~\bibnamefont{Wahl}}, \bibnamefont{and}
  \bibinfo{author}{\bibfnamefont{T.}~\bibnamefont{Martin}},
  \bibinfo{journal}{Phys. Rev. B} \textbf{\bibinfo{volume}{86}},
  \bibinfo{pages}{125425} (\bibinfo{year}{2012}).

\bibitem[{\citenamefont{Levkivskyi and Sukhorukov}(2008)}]{MZI_nu2}
\bibinfo{author}{\bibfnamefont{I.~P.} \bibnamefont{Levkivskyi}}
  \bibnamefont{and} \bibinfo{author}{\bibfnamefont{E.~V.}
  \bibnamefont{Sukhorukov}}, \bibinfo{journal}{Phys. Rev. B}
  \textbf{\bibinfo{volume}{78}}, \bibinfo{pages}{045322}
  (\bibinfo{year}{2008}).

\bibitem[{\citenamefont{Neder}(2012)}]{Neder_fractionalization}
\bibinfo{author}{\bibfnamefont{I.}~\bibnamefont{Neder}},
  \bibinfo{journal}{Phys. Rev. Lett.} \textbf{\bibinfo{volume}{108}},
  \bibinfo{pages}{186404} (\bibinfo{year}{2012}).

\bibitem[{\citenamefont{Berg et~al.}(2009)\citenamefont{Berg, Oreg, Kim, and
  von Oppen}}]{Berg_fractional_charges}
\bibinfo{author}{\bibfnamefont{E.}~\bibnamefont{Berg}},
  \bibinfo{author}{\bibfnamefont{Y.}~\bibnamefont{Oreg}},
  \bibinfo{author}{\bibfnamefont{E.-A.} \bibnamefont{Kim}}, \bibnamefont{and}
  \bibinfo{author}{\bibfnamefont{F.}~\bibnamefont{von Oppen}},
  \bibinfo{journal}{Phys. Rev. Lett.} \textbf{\bibinfo{volume}{102}},
  \bibinfo{pages}{236402} (\bibinfo{year}{2009}).

\bibitem[{\citenamefont{Levkivskyi and
  Sukhorukov}(2012)}]{shot_noise_thermometry}
\bibinfo{author}{\bibfnamefont{I.~P.} \bibnamefont{Levkivskyi}}
  \bibnamefont{and} \bibinfo{author}{\bibfnamefont{E.~V.}
  \bibnamefont{Sukhorukov}}, \bibinfo{journal}{Phys. Rev. Lett.}
  \textbf{\bibinfo{volume}{109}}, \bibinfo{pages}{246806}
  (\bibinfo{year}{2012}).

\bibitem[{\citenamefont{Kovrizhin and Chalker}(2011)}]{KovrizhinChalker}
\bibinfo{author}{\bibfnamefont{D.~L.} \bibnamefont{Kovrizhin}}
  \bibnamefont{and} \bibinfo{author}{\bibfnamefont{J.~T.}
  \bibnamefont{Chalker}}, \bibinfo{journal}{Phys. Rev. B}
  \textbf{\bibinfo{volume}{84}}, \bibinfo{pages}{085105}
  (\bibinfo{year}{2011}).

\bibitem[{\citenamefont{Milletar{\`i} and Rosenow}(2013)}]{shot_noise_nu2}
\bibinfo{author}{\bibfnamefont{M.}~\bibnamefont{Milletar{\`i}}}
  \bibnamefont{and} \bibinfo{author}{\bibfnamefont{B.}~\bibnamefont{Rosenow}},
  \bibinfo{journal}{Phys. Rev. Lett.} \textbf{\bibinfo{volume}{111}},
  \bibinfo{pages}{136807} (\bibinfo{year}{2013}).

\bibitem[{\citenamefont{Altimiras et~al.}(2009)\citenamefont{Altimiras,
  le~Sueur, Gennser, Cavanna., Mailly, and Pierre}}]{Altimiras_spectroscopy}
\bibinfo{author}{\bibfnamefont{C.}~\bibnamefont{Altimiras}},
  \bibinfo{author}{\bibfnamefont{H.}~\bibnamefont{le~Sueur}},
  \bibinfo{author}{\bibfnamefont{U.}~\bibnamefont{Gennser}},
  \bibinfo{author}{\bibfnamefont{A.}~\bibnamefont{Cavanna.}},
  \bibinfo{author}{\bibfnamefont{D.}~\bibnamefont{Mailly}}, \bibnamefont{and}
  \bibinfo{author}{\bibfnamefont{F.}~\bibnamefont{Pierre}},
  \bibinfo{journal}{Nat. Phys.} \textbf{\bibinfo{volume}{6}},
  \bibinfo{pages}{34} (\bibinfo{year}{2009}).

\bibitem[{\citenamefont{le~Sueur et~al.}(2010)\citenamefont{le~Sueur,
  Altimiras, Gennser, Cavanna, Mailly, and Pierre}}]{leSueur_energy_relaxation}
\bibinfo{author}{\bibfnamefont{H.}~\bibnamefont{le~Sueur}},
  \bibinfo{author}{\bibfnamefont{C.}~\bibnamefont{Altimiras}},
  \bibinfo{author}{\bibfnamefont{U.}~\bibnamefont{Gennser}},
  \bibinfo{author}{\bibfnamefont{A.}~\bibnamefont{Cavanna}},
  \bibinfo{author}{\bibfnamefont{D.}~\bibnamefont{Mailly}}, \bibnamefont{and}
  \bibinfo{author}{\bibfnamefont{F.}~\bibnamefont{Pierre}},
  \bibinfo{journal}{Phys. Rev. Lett.} \textbf{\bibinfo{volume}{105}},
  \bibinfo{pages}{056803} (\bibinfo{year}{2010}).

\bibitem[{\citenamefont{Lunde et~al.}(2010)\citenamefont{Lunde, Nigg, and
  B{\"u}ttiker}}]{EC_equilibration}
\bibinfo{author}{\bibfnamefont{A.~M.} \bibnamefont{Lunde}},
  \bibinfo{author}{\bibfnamefont{S.~E.} \bibnamefont{Nigg}}, \bibnamefont{and}
  \bibinfo{author}{\bibfnamefont{M.}~\bibnamefont{B{\"u}ttiker}},
  \bibinfo{journal}{Phys. Rev. B} \textbf{\bibinfo{volume}{81}},
  \bibinfo{pages}{041311} (\bibinfo{year}{2010}).

\bibitem[{\citenamefont{Degiovanni et~al.}(2010)\citenamefont{Degiovanni,
  Grenier, F{\`e}ve, Altimiras, le~Sueur, and
  Pierre}}]{degiovanni_plasmon_2010}
\bibinfo{author}{\bibfnamefont{P.}~\bibnamefont{Degiovanni}},
  \bibinfo{author}{\bibfnamefont{C.}~\bibnamefont{Grenier}},
  \bibinfo{author}{\bibfnamefont{G.}~\bibnamefont{F{\`e}ve}},
  \bibinfo{author}{\bibfnamefont{C.}~\bibnamefont{Altimiras}},
  \bibinfo{author}{\bibfnamefont{H.}~\bibnamefont{le~Sueur}}, \bibnamefont{and}
  \bibinfo{author}{\bibfnamefont{F.}~\bibnamefont{Pierre}},
  \bibinfo{journal}{Phys. Rev. B} \textbf{\bibinfo{volume}{81}},
  \bibinfo{pages}{121302} (\bibinfo{year}{2010}).

\bibitem[{\citenamefont{von Delft and
  Schoeller}(1998)}]{bosonization_refermionization}
\bibinfo{author}{\bibfnamefont{J.}~\bibnamefont{von Delft}} \bibnamefont{and}
  \bibinfo{author}{\bibfnamefont{H.}~\bibnamefont{Schoeller}},
  \bibinfo{journal}{Ann. Phys. (Berlin)} \textbf{\bibinfo{volume}{7}},
  \bibinfo{pages}{225} (\bibinfo{year}{1998}).

\bibitem[{\citenamefont{Bocquillon
  et~al.}(2013{\natexlab{b}})\citenamefont{Bocquillon, Freulon, Berroir,
  Degiovanni, Pla\c{c}ais, Cavanna, Jin, and F\`eve}}]{Bocquillon_interedge}
\bibinfo{author}{\bibfnamefont{E.}~\bibnamefont{Bocquillon}},
  \bibinfo{author}{\bibfnamefont{V.}~\bibnamefont{Freulon}},
  \bibinfo{author}{\bibfnamefont{J.-M.} \bibnamefont{Berroir}},
  \bibinfo{author}{\bibfnamefont{P.}~\bibnamefont{Degiovanni}},
  \bibinfo{author}{\bibfnamefont{B.}~\bibnamefont{Pla\c{c}ais}},
  \bibinfo{author}{\bibfnamefont{A.}~\bibnamefont{Cavanna}},
  \bibinfo{author}{\bibfnamefont{Y.}~\bibnamefont{Jin}}, \bibnamefont{and}
  \bibinfo{author}{\bibfnamefont{G.}~\bibnamefont{F\`eve}},
  \bibinfo{journal}{Nat. Commun.} \textbf{\bibinfo{volume}{4}},
  \bibinfo{pages}{1839} (\bibinfo{year}{2013}{\natexlab{b}}).

\bibitem[{\citenamefont{Martin}(2004)}]{Martin_Houches}
\bibinfo{author}{\bibfnamefont{T.}~\bibnamefont{Martin}}, in
  \emph{\bibinfo{booktitle}{Nanophysics: Coherence and Transport: \'Ecole
  d'\'Et\'e de Physique des Houches: Session LXXXI: 28 June-30 July, 2004: Euro
  Summer School, Nato Advanced Study Institute, \'Ecole Th\'ematique du CNRS}},
  edited by \bibinfo{editor}{\bibfnamefont{H.}~\bibnamefont{Bouchiat}},
  \bibinfo{editor}{\bibfnamefont{Y.}~\bibnamefont{Gefen}},
  \bibinfo{editor}{\bibfnamefont{S.}~\bibnamefont{Gu{\'e}ron}},
  \bibinfo{editor}{\bibfnamefont{G.}~\bibnamefont{Montambaux}},
  \bibnamefont{and} \bibinfo{editor}{\bibfnamefont{J.}~\bibnamefont{Dalibard}}
  (\bibinfo{publisher}{Elsevier, Amsterdam}, \bibinfo{year}{2004}), p.
  \bibinfo{pages}{283}.

\bibitem[{\citenamefont{Martin and Landauer}(1992)}]{Martin_1992}
\bibinfo{author}{\bibfnamefont{T.}~\bibnamefont{Martin}} \bibnamefont{and}
  \bibinfo{author}{\bibfnamefont{R.}~\bibnamefont{Landauer}},
  \bibinfo{journal}{Phys. Rev. B} \textbf{\bibinfo{volume}{45}},
  \bibinfo{pages}{1742} (\bibinfo{year}{1992}).

\bibitem[{SM()}]{SM}
\bibinfo{note}{See Suplemental Material at [url] for the justification of the
  scattering approach used to describe the QPC, a brief discussion of zero
  temperature results, and details on the numerical integration.}

\bibitem[{\citenamefont{Grenier et~al.}(2011)\citenamefont{Grenier, Herv{\'e},
  Bocquillon, Parmentier, Pla\c{c}ais, Berroir, F{\`e}ve, and
  Degiovanni}}]{SES_tomography}
\bibinfo{author}{\bibfnamefont{C.}~\bibnamefont{Grenier}},
  \bibinfo{author}{\bibfnamefont{R.}~\bibnamefont{Herv{\'e}}},
  \bibinfo{author}{\bibfnamefont{E.}~\bibnamefont{Bocquillon}},
  \bibinfo{author}{\bibfnamefont{F.}~\bibnamefont{Parmentier}},
  \bibinfo{author}{\bibfnamefont{B.}~\bibnamefont{Pla\c{c}ais}},
  \bibinfo{author}{\bibfnamefont{J.-M.} \bibnamefont{Berroir}},
  \bibinfo{author}{\bibfnamefont{G.}~\bibnamefont{F{\`e}ve}}, \bibnamefont{and}
  \bibinfo{author}{\bibfnamefont{P.}~\bibnamefont{Degiovanni}},
  \bibinfo{journal}{New J. of Phys.} \textbf{\bibinfo{volume}{13}},
  \bibinfo{pages}{093007} (\bibinfo{year}{2011}).

\bibitem[{\citenamefont{Hahn}(2005)}]{Cuba}
\bibinfo{author}{\bibfnamefont{T.}~\bibnamefont{Hahn}},
  \bibinfo{journal}{Comput. Phys. Commun.} \textbf{\bibinfo{volume}{168}},
  \bibinfo{pages}{78} (\bibinfo{year}{2005}).

\bibitem[{\citenamefont{Degiovanni et~al.}(2009)\citenamefont{Degiovanni,
  Grenier, and F{\`e}ve}}]{Degiovanni_decoherence}
\bibinfo{author}{\bibfnamefont{P.}~\bibnamefont{Degiovanni}},
  \bibinfo{author}{\bibfnamefont{C.}~\bibnamefont{Grenier}}, \bibnamefont{and}
  \bibinfo{author}{\bibfnamefont{G.}~\bibnamefont{F{\`e}ve}},
  \bibinfo{journal}{Phys. Rev. B} \textbf{\bibinfo{volume}{80}},
  \bibinfo{pages}{241307} (\bibinfo{year}{2009}).

\bibitem[{\citenamefont{Lee and Levitov}()}]{Lee_orthogonality_catastrophe}
\bibinfo{author}{\bibfnamefont{H.}~\bibnamefont{Lee}} \bibnamefont{and}
  \bibinfo{author}{\bibfnamefont{L.}~\bibnamefont{Levitov}},
  \bibinfo{note}{arXiv:cond-mat/9312013}.

\bibitem[{\citenamefont{Bocquillon et~al.}(2012)\citenamefont{Bocquillon,
  Parmentier, Grenier, Berroir, Degiovanni, Glattli, Pla\c{c}ais, Cavanna, Jin,
  and F{\`e}ve}}]{SES_HBT}
\bibinfo{author}{\bibfnamefont{E.}~\bibnamefont{Bocquillon}},
  \bibinfo{author}{\bibfnamefont{F.-D.} \bibnamefont{Parmentier}},
  \bibinfo{author}{\bibfnamefont{C.}~\bibnamefont{Grenier}},
  \bibinfo{author}{\bibfnamefont{J.-M.} \bibnamefont{Berroir}},
  \bibinfo{author}{\bibfnamefont{P.}~\bibnamefont{Degiovanni}},
  \bibinfo{author}{\bibfnamefont{D.-C.} \bibnamefont{Glattli}},
  \bibinfo{author}{\bibfnamefont{B.}~\bibnamefont{Pla\c{c}ais}},
  \bibinfo{author}{\bibfnamefont{A.}~\bibnamefont{Cavanna}},
  \bibinfo{author}{\bibfnamefont{Y.}~\bibnamefont{Jin}}, \bibnamefont{and}
  \bibinfo{author}{\bibfnamefont{G.}~\bibnamefont{F{\`e}ve}},
  \bibinfo{journal}{Phys. Rev. Lett.} \textbf{\bibinfo{volume}{108}},
  \bibinfo{pages}{196803} (\bibinfo{year}{2012}).

\bibitem[{\citenamefont{F{\`e}ve et~al.}(2008)\citenamefont{F{\`e}ve,
  Degiovanni, and Jolicoeur}}]{Feve_flying_qubits}
\bibinfo{author}{\bibfnamefont{G.}~\bibnamefont{F{\`e}ve}},
  \bibinfo{author}{\bibfnamefont{P.}~\bibnamefont{Degiovanni}},
  \bibnamefont{and}
  \bibinfo{author}{\bibfnamefont{T.}~\bibnamefont{Jolicoeur}},
  \bibinfo{journal}{Phys. Rev. B} \textbf{\bibinfo{volume}{77}},
  \bibinfo{pages}{035308} (\bibinfo{year}{2008}).

\bibitem[{\citenamefont{Bocquillon}(2012)}]{these_Bocquillon}
\bibinfo{author}{\bibfnamefont{E.}~\bibnamefont{Bocquillon}}, Ph.D. thesis,
  \bibinfo{school}{Universit{\'e} Paris VI} (\bibinfo{year}{2012}).

\bibitem[{\citenamefont{Dubois et~al.}(2013)\citenamefont{Dubois, Jullien,
  Portier, Roche, Cavanna., Jin, Wegscheider, Roulleau, and
  Glattli}}]{Dubois_2013}
\bibinfo{author}{\bibfnamefont{J.}~\bibnamefont{Dubois}},
  \bibinfo{author}{\bibfnamefont{T.}~\bibnamefont{Jullien}},
  \bibinfo{author}{\bibfnamefont{F.}~\bibnamefont{Portier}},
  \bibinfo{author}{\bibfnamefont{P.}~\bibnamefont{Roche}},
  \bibinfo{author}{\bibfnamefont{A.}~\bibnamefont{Cavanna.}},
  \bibinfo{author}{\bibfnamefont{Y.}~\bibnamefont{Jin}},
  \bibinfo{author}{\bibfnamefont{W.}~\bibnamefont{Wegscheider}},
  \bibinfo{author}{\bibfnamefont{P.}~\bibnamefont{Roulleau}}, \bibnamefont{and}
  \bibinfo{author}{\bibfnamefont{D.~C.} \bibnamefont{Glattli}},
  \bibinfo{journal}{Nature (London)} \textbf{\bibinfo{volume}{502}},
  \bibinfo{pages}{659} (\bibinfo{year}{2013}).

\end{thebibliography}

\end{document}